\documentclass{desyproc}

\begin{document}
\title{Detecting an infrared Photon within an Hour --
Transition-Edge Detector at ALPS-II}

\author{{\slshape Jan Dreyling-Eschweiler$^{1,2}$, Dieter Horns$^{2}$, Friederike Januschek$^{1}$ and Axel Lindner$^{1}$, for the ALPS-II collaboration}\\[1ex]
$^1$Deutsches Elektronen-Synchrotron (DESY), Hamburg, Germany\\
$^2$University of Hamburg, Hamburg, Germany}

\contribID{dreyling-eschweiler\_jan}

\desyproc{DESY-PROC-2013-XX}
\acronym{Patras 2013} 
\doi  

\maketitle


\begin{abstract} 
An essential design requirement of the ALPS-II experiment is the efficient detection of single photons with a very low instrumental background of 10~$\mu$Hz.  
In 2011 the ALPS collaboration started to set up a TES detector (Transition-Edge Sensor) for ALPS-II, the second phase of the experiment. 
Since mid of 2013 the setup is ready for characterization in the ALPS laboratory: an ADR cryostat (Adiabatic Demagnetization Refrigerator) as millikelvin environment, a low noise SQUID (Superconducting Quantum Interference Device) with electronics for read-out and a fiber-coupled high-efficient TES for near-infrared photons as sensor. 
First measurements have shown a good discrimination between noise and 1064~nm signals. 
\end{abstract}


\section{Photon detection at ALPS-I and ALPS-II} 
The ALPS-I experiment has provided the most constraining limits for photon-ALP coupling for a light-shining-through-a-wall experiment~\cite{Ehret:2010mh}.
The detector was a CCD camera (Charged-coupled Device) with a quantum efficiency $>$90~\% for the ALPS-I wavelength of 532~nm and a dark current of about 0.0008~$e$~s$^{-1}$ per pixel.   
Data acquisition was done with 1~h data respectively dark frames being in the linear noise regime where the dark current dominates the read-out noise, but limited by charged particle background like cosmics or decay products. For 1~h frames the overall detector noise is about 0.0018 s$^{-1}$ including the read-out noise of the CCD and a beam focus on 3x3 pixel~\cite{Bahre:2013ywa}.

In the second phase, ALPS-II, the overall sensitivity of the experiment will mainly be improved by higher laser power, a regeneration cavity behind the wall and a length up to 200~m~\cite{Bahre:2013ywa, Doebrich:2013}.  
But by switching to a laser wavelength of 1064~nm the quantum efficiency of the CCD drops below 1.5~\%~\cite{Seggern:2013} because of the Si band gap.  
So in parallel the ALPS collaboration is looking for an alternative detector. A promising candidate is a TES having no dark counts intrinsically~\cite{Cabrera:1998} and providing an energy and time resolution in addition compared to a CCD. A quantum efficiency for near-infrared photons near unity has been realized~\cite{Lita:2010}. 


\section{TES detectors: working principle and realizations}


A TES is operating as a microcalorimeter: The sensor consists of a film that is biased by an electrical current into the superconducting phase transition. If energy is deposited e.g. by a photon, the TES heats up fast and cools down slowly because the TES is weakly linked to the cold bath and relaxing to its working point~\cite{Irwin:2005}. 
The change of temperature results in a change of resistance and, in a voltage-biased circuit, in a change of current which can be measured by an inductive-coupled and impedance-matched SQUID and read out as a voltage change with proper electronics.
In the linear description of these electro-thermal system the integral of the pulse is proportional to the energy input. 
 

The realized TES detectors reached a big bandwidth in the last 20 years: They cover the electromagnetic spectrum from gamma rays, over X-rays and the optical/infrared regime, through millimeter range. Applications are found in spectroscopy, astronomy or direct Dark Matter searches for example. For ALPS there is an overlap with the field of quantum information, which uses TES detectors as single-photon counter at the telecommunication wavelength 1330/1550~nm. 
The research and development of fiber-coupled high-efficient TES for detection of near-infrared photons is actively carried out at NIST (National Institute of Standards and Technology) in the U.S. and AIST (National Institute of Advanced Industrial Science and Technology) in Japan. Both metrology institutes reached a near unity efficiency for detecting single infrared photons~\cite{Lita:2010, Fukuda:2011} fitting to ALPS detection requirements. For these devices time resolution is up to $\sim$1~$\mu$s and energy resolution $\sim$0.1~eV. The superconducting transition of the sensor material is about $\sim$140~mK for the W-based TESs of NIST and about $\sim$300~mK for the Ti/Au-based TESs of AIST.

 
\section{Realization for ALPS-II: history, croyostat and sensors}


The ALPS collaboration set to work on TESs, SQUIDs and mK-cryogenics in the end of 2010. 
The primary goal has been to operate and characterize a TES. A focus has been on the background.
Only upper limits have been set by previous studies~\cite{Miller:2003, Miller:2007}.
In early 2011 we had the opportunity to see the operation of SQUIDs coupled to NIST TES assembled in a DR (Dilution Refrigerator) at the PTB (Physikalisch-Technische Bundesanstalt) in Berlin. 
We tried to establish a TES setup at the University of Camerino, Italy, during two measurement periods in 2011, which were limited due the evaporating liquid Helium as pre-cooling technique~\cite{Cantatore:2011}: In a dip-in DR we assembled a low-efficiency TES chip from INRIM (L'Istituto Nazionale di Ricerca Metrologica) coupled to a SQUID from the company Magnicon, Germany. The optical fiber fed in the cryostat wasn't directly coupled to the TES but its end pointed to the sensor area. 
We succesfully achieved single photon detection with this first setup~\cite{Cantatore:2013}.
Since the end of 2012 we operate and characterize an ADR cryostat from the company Entropy, Germany. First time we operated it at PTB, Berlin, for a good knowledge transfer. 
There we used a sensor module equipped with PTB SQUIDs and NIST TESs as proof of principle. 
After moving the cryostat to Hamburg in the end of 2012, in early 2013 we started operating sensor modules within the ADR in the ALPS laboratory.



\begin{figure}[h]
  \begin{minipage}{0.45\textwidth}
\centerline{\includegraphics[width=0.75\textwidth]{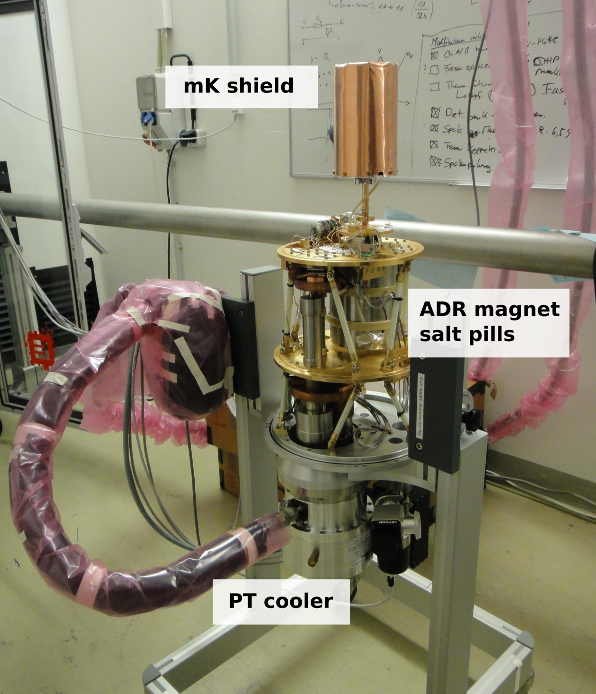}}
\caption{Open ADR cryostat (upside down) with different cooling stages in the ALPS-IIa lab: Here at the top a mK copper shield is connected to the cold finger where the sensors are located inside.}
\label{Fig:ADR}
  \end{minipage}
  \hspace{0.1\textwidth}
  \begin{minipage}{0.45\textwidth}
\centerline{\includegraphics[width=0.8\textwidth]{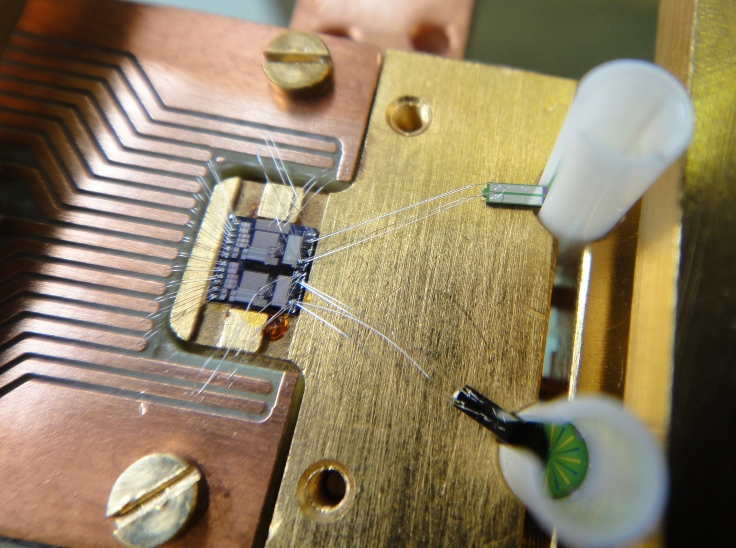}}
\caption{NIST module with two channels: Left at the end of the PCB the SQUID chip is located. Bondwires connect the TES, which has a shape similiar to a table-tennis bat. Around the chip is a ceramic split sleeve to connect a fiber with a ferrule end of a common FC connector. The sensitive area of doped tungsten (W) is about 25x25~{$\mu$}m.}
\label{Fig:module}
  \end{minipage}
  \\
  \center
  \begin{minipage}{0.5\textwidth}
\centerline{\includegraphics[width=0.9\textwidth]{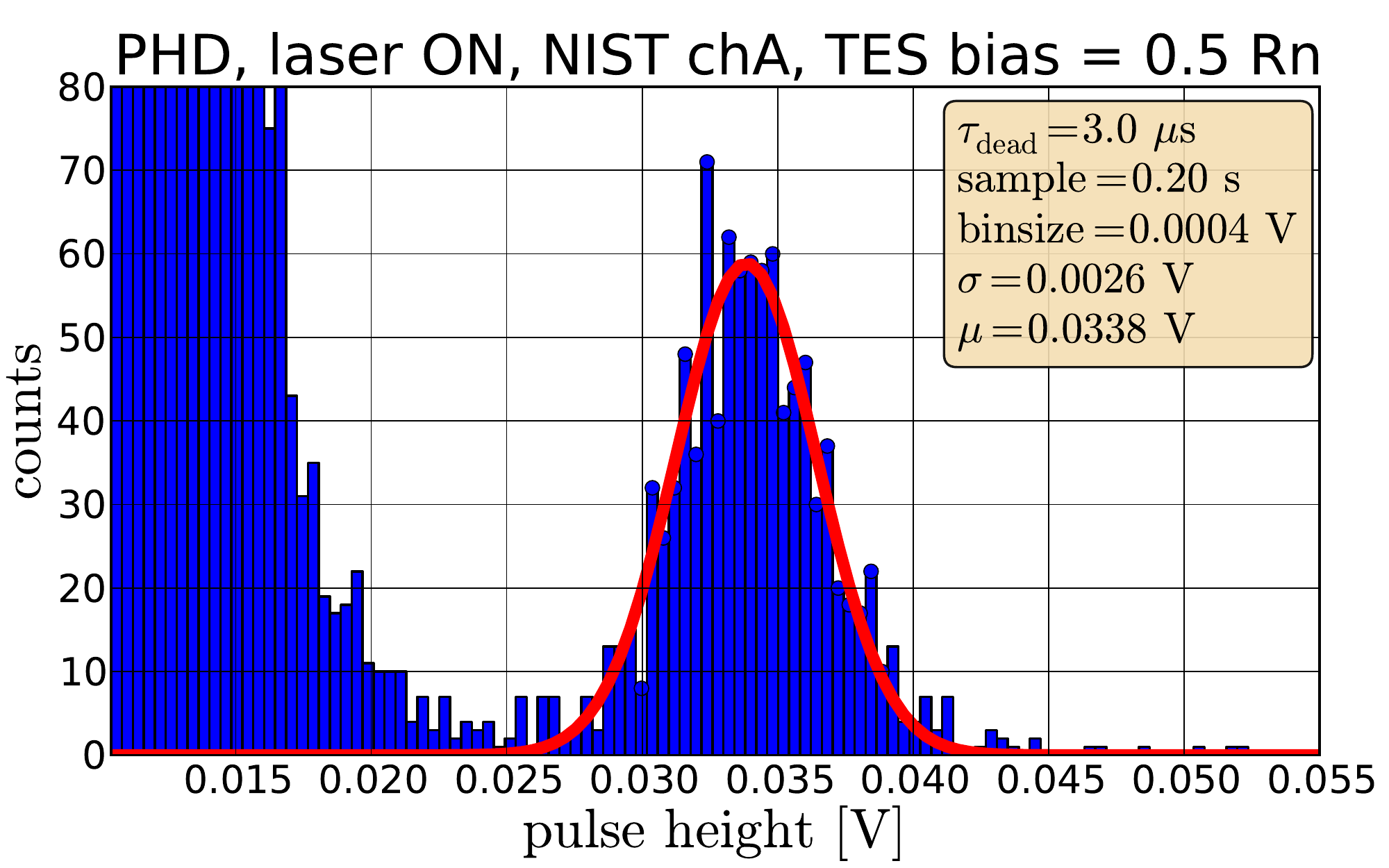}}
\caption{Pulse height distribution with a signal peak (red Gauss shape) of 1066.7~nm photons. Noise counts are below $\sim$ 25 mV.}
\label{Fig:phd}
  \end{minipage}
\end{figure}

The ADR is a no-liquid-cryogens cryostat with a closed pre-cooling He cycle: integrated is a two-stage pulse-tube cooler with which the 4~K stage is established, see Fig.~\ref{Fig:ADR}. 
Attached to that is a superconducting magnet\footnote{With 40~A current a magnetic field of 6~T is realized.} which surrounds a double-stage salt pill unit which can be coupled/decoupled to the 4~K stage by a piezo-driven motor.
An adiabatic demagnetization cycle reaches 30~mK as lowest temperature after $>$90~min. 
By regulating the magnet a constant bath temperature for sensors is achieved: For example the hold time for 80~mK~$\pm$ 25~$\mu$K (rms) is about 24~h. 
The remnant magnetic field for regulating is screened by a cryoperm layer around the magnet passively and doesn't affect the operation of the sensors.


ALPS has two sensor modules, each with two channels and with optimized TESs for 1064~nm: One with TESs from AIST, where the single mode fiber is glued to the sensitive area~\cite{Fukuda:2011}, a second with TESs from NIST, where the single mode fiber is connected with the standard way of FC connectors~\cite{Miller:2012}, see Fig.~\ref{Fig:module}. 
Both sensor modules are connected to PTB dc 2-stage SQUIDs, which were developed for low-noise TES readout. 
With a readout electronic (XXF-1) from the company Magnicon the SQUID and TES sensors are set to the working point. 
For the first measurements the data acquisition was done with an oscilloscope (DPO700c from Tektronix).


\section{First results}


We successfully set up the cryogenic mK-environment with an ADR cryostat in the ALPS laboratory. In several cool downs we operate the sensor modules as a single photon detector for the ALPS-II wavelength.

As a first important result for ALPS-II, signal and noise (electronic, Johnson and thermal noise) are distinguishable, see fig.~\ref{Fig:phd}. 
 In this measurement we set the sensor module in an arbitrarily chosen working point and realized a single photon rate with an attenuated laser (1066.7~nm) as a signal. The relative energy resolution is $\Delta E/E = 7.7\,\%$.

Further measurements for optimization of the working point and long time measurements for background analysis are on the way~\cite{Dreyling:2014}. 
Thermal photons of 300~K were found to be one main component for background events \cite{Miller:2007}.


\section*{Acknowledgments}

The ALPS collaboration wants to thank everybody who helped along the way to setup this TES detector. Special thanks go to Joern Beyer (PTB), Daiji Fukuda (AIST) and Sae Woo Nam (NIST) for supplying the superconducting devices.


\begin{footnotesize}

\end{footnotesize}


\end{document}